# Effects resulting from magnetic interactions in low-dimensional systems

José Holanda[1,2,3,4,5*]

[1]Programa de Pós-Graduação em Engenharia Física, Universidade Federal Rural de Pernambuco, 54518-430, Cabo de Santo Agostinho, Pernambuco, Brazil
[2]Group of Optoelectronics and Spintronics, Universidade Federal Rural de Pernambuco, 54518-430, Cabo de Santo Agostinho, Pernambuco, Brazil
[3]Unidade Acadêmica do Cabo de Santo Agostinho, Universidade Federal Rural de Pernambuco, 54518-430, Cabo de Santo Agostinho, Pernambuco, Brazil
[4]Programa de Pós-Graduação em Física Aplicada, Universidade Federal Rural de Pernambuco, 52171-900, Recife, Pernambuco, Brazil
[5]Programa de Pós Graduação em Tecnologias Energéticas e Nucleares (Proten), Universidade Federal de Pernambuco, Recife, 50740-545, PE, Brazil

**Abstract**

This research delves into the critical effects of magnetic interactions in low-dimensional systems, offering invaluable insights that deepen our comprehension of magnetic behavior at the nanoscale. By implementing this innovative approach, one can unequivocally identify two distinct magnetic states: demagnetizing and magnetizing. The resulting measurements significantly enhance our grasp of the magnetic dynamics within these nanostructures, paving the way for spin-wave excitations. To validate the effectiveness of this methodology, it was conducted rigorous numerical simulations on a diverse array of nanostructures, including one-dimensional nanowires and three-dimensional hexagonal arrays of nanowires. Each nanowire is precisely modeled as a chain of interacting ellipsoidal grains, illustrating the intricate nature of these magnetic interactions.

*Corresponding author: joseholanda.silvajunior@ufrpe.br
Orcid ID: https://orcid.org/0000-0002-8823-368X

# I. INTRODUCTION

Recent investigations underscore the growing significance of controlling the size, shape, and orientation of one-dimensional magnetic nanostructures, particularly nanowires [1-8]. These magnetic nanostructures have garnered immense attention due to their transformative applications in advanced spintronics [9], cutting-edge sensor devices, and innovative biomedical technologies [10, 11]. They represent a crucial research frontier for exploring diverse magnetic phenomena, including magnetization reversal and complex magnetic-optic processes [1-11]. The pursuit of tunable magnetic interactions through novel methodologies promises to yield groundbreaking advancements in this field. The fabrication of nanostructures with well-defined dimensions benefits immensely from template-assisted growth techniques. Notably, the electrochemical synthesis of metallic magnetic nanowires has emerged as a highly effective approach for producing nanoscaled systems [12-19]. Electrodeposition within alumina membranes stands out as the most prevalent method for manufacturing these systems, resulting in highly ordered arrays of magnetic nanowires that are intricately embedded within the membrane pores [12-16]. Noteworthy studies reveal the presence of various magnetization reversal modes, with the coherent configuration often dominating [1-17].

Understanding the effects of magnetic interactions in particulate media is critical, as these interactions can profoundly influence the magnetic properties of nanostructures [20-24]. Generally, the magnetization resulting from an applied magnetic field comprises both reversible and irreversible components. This complexity renders it impossible to disentangle these components in a hysteresis curve, a fact widely acknowledged in the field. As a result, much vital information remains concealed, leading to a limited grasp of remanent measures. The exploration of the remanent state serves as a powerful tool for investigating magnetic interactions. In specific systems, established delta-m ($\Delta m$) curves provide compelling insights, allowing for meaningful comparisons between isothermal remanent magnetization (IRM) and direct current demagnetization (DCD) data [20, 25-30].

In an IRM(H) experiment, the investigation begins with a demagnetized sample that has been meticulously cooled in a zero magnetic field. Once the desired temperature is attained, a small magnetic field is applied, allowing sufficient time for the system to reach thermodynamic equilibrium. Following the removal of this magnetic field, the remanence is carefully measured. This procedure is systematically repeated, progressively increasing the magnetic field strength until the sample reaches saturation, at which point the remanence achieves its peak value. In contrast, the DCD(H) experiment commences with the sample already in a saturated state. At a fixed temperature, the applied magnetic field is gradually inverted until it achieves a small value that opposes the initial magnetization. After this inversion, the magnetic field is switched off, and the remanence is measured once more. This cycle is repeated, incrementally increasing the strength of the inverted magnetic field until saturation is reached. Through these carefully designed processes, one can qualitatively assess the predominant types of magnetic interactions - be they magnetizing or demagnetizing - within various magnetic samples [20, 25-34]. In the absence of such interactions, the magnetic system adheres to the principles outlined by the Stoner and Wohlfarth model [20, 30].

This study presents a novel approach to accurately determine the intensity of predominant magnetic interactions in nanostructures. Furthermore, it is conducted a comprehensive numerical analysis to explore how the intensity of these magnetic interactions varies with the angle of the applied magnetic field. These findings illuminate the behavior of these interactions based on their intensity, unveiling two distinctive states: magnetizing and demagnetizing. This research not only enhances our understanding of magnetic interactions in nanostructures but also paves the way for advancements in applications leveraging these fundamental principles.

**II. APPROACH FOR THE NUMERICAL CALCULATIONS**

The interactions within magnetic nanostructures have been qualitatively examined through delta-m ($\Delta m$) curves [20, 25-30]. These delta-m curves provide a critical comparison between the IRM(H) and the DCD(H). Both IRM(H) and DCD(H) are measured using consistent experimental methodologies, differing only in the magnetic states of the sample at the start of each measurement [16]. The foundational work of Stoner and Wohlfarth [20, 30] establishes a vital relationship between IRM(H) and DCD(H) for non-interacting particles, formulated as follows:

$$m_{d1} = 1 - 2m_r, \qquad (1)$$

where the variables $m_{d1} = DCD(H)_1/IRM(H_{Max})$ and $m_r = IRM(H)/IRM(H_{Max})$ correspond to specific characteristics of the magnetic materials. In contrast, Henkel's theory $(m_d(H) \times m_r(H))$ articulates that any deviation from this relationship in real-world systems arises from interactions between nanoparticles [25-29]. This insight into Henkel curves underscores that when magnetic interactions are present, experimental data inherently diverges from the predictions made by Stoner and Wohlfarth's equation. Such divergence is crucial since the original equation presumes an equivalency between the magnetizing and demagnetizing processes. To precisely assess the nature of these interactions, a pivotal term, delta-m ($\Delta m$), is introduced within the framework of the original equation [25-29]. The interpretation of $\Delta m$ is clear:

$$m_{d2}(H) = \Delta m + [1 - 2m_r(H)], \qquad (2)$$

where $m_{d2} = DCD(H)_2/IRM(H_{Max})$. Now, if $\Delta m < 0$, the predominant interactions are demagnetizing (PID); conversely, if $\Delta m > 0$, the predominant interactions are magnetizing (PIM). Understanding these interactions is not merely an academic exercise; it is fundamentally important for advancing the field. This proposition is that the area defined by the equations serves as a quantitative measure of the strength of these magnetic interactions, and are very important for experimental and numerical analyses. By employing Riemann's sum to redefine the originals equations (equations (1) and (2)), one can derive a more robust understanding of magnetic interactions within these nanostructures:

$$I_1 = \lim_{\|\delta\| \to 0} \sum_{i=1}^{N} m_{d1}^i \, \delta_i m_r. \qquad (3)$$

Using equation (2), one can derive the following expression:

$$I_2 = \lim_{\|\delta\| \to 0} \sum_{i=1}^{N} m_{d2}^i \, \delta_i m_r, \qquad (4)$$

where, $\|\delta\| = \text{Max}\{\delta_i m_r, 1 \le i \le N\}$. Next, the intensity of the magnetic interactions is determined using expressions (3) and (4) as follows:

$$I = \lim_{\|\delta\| \to 0} \sum_{i=1}^{N} \left| m_{d2}^i - m_{d1}^i \right| \delta_i m_r = \lim_{\|\delta\| \to 0} \sum_{i=1}^{N} |\Delta m_i| \delta_i m_r, \qquad (5)$$

with $m_r$ in the specified interval $\left[m_r^{in}, m_r^{fi}\right]$. Given that expression (5) can represent both PID and PIM interactions, one can formulate the following relationships:

$$I = \lim_{\|\delta\| \to 0} \sum_{i=1}^{N} |\Delta m_i| \delta_i m_r \to \begin{cases} \Delta m_i < 0 \text{ for PID}, \\ \Delta m_i > 0 \text{ for PIM}. \end{cases} \qquad (6)$$

## III. RESULTS AND DISCUSSION

The numerical results derived from this approach focused on two distinct nanostructures across two dimensions: a) 1D, featuring a single nanowire, and b) 3D, comprising a hexagonal array of nanowires. Each nanowire is effectively represented by a chain of interacting ellipsoidal grains, highlighting the intricate relationships within these low-dimensional systems.

### a) 1D, featuring a single nanowire

**Figure 1(a)** presents a compelling expression for magnetic energy density in the presence of an applied magnetic field, as referenced in sources [2, 16]. This relationship underscores the significant impact of external fields on energy dynamics:

$$E_{NW} = \pi M_{Max}^2 \left[ \left( \frac{3V_e k_n}{\pi r^3} \right) + (1 - 3N_{\parallel}) \right] \sin^2(\varphi + \theta_H) - M_{Max} H \cos(\varphi). \tag{7}$$

In equation (7), the first term represents the energy density associated with the dipolar interactions between grains, the second term refers to the demagnetizing energy density, and the last term represents the Zeeman energy density. Additionally, $\varphi = \theta - \theta_H$ is the angle between the magnetization and the magnetic field. Here, $M_{Max}$ denotes the maximum magnetization in the presence of a magnetic field, r is the center-to-center distance between grains, $N_{\parallel}$ is the demagnetizing factor corresponding to the long axis of the grains in the chain, $V_e$ is the volume of each grain, and $k_n$ is defined as the summation $k_n = \sum_{j=1}^{n}(n-j)/nj^3$, which includes combinations for magnetic inter-grain interactions. In this summation, n represents the total number of grains forming the nanowire, while j denotes the number of grain pairs that are equidistant. It is possible to rewrite equation (7) as follows:

$$\varepsilon_{NW} = \frac{1}{2} \sin^2(\varphi + \theta_H) - f\cos(\varphi). \tag{8}$$

The factor $f$ is defined as $f = H/H_{eff}$, where $H_{eff} = 2\pi M_{Max} \left[ 1 + 3(V_e k_n/\pi r^3 - N_{\parallel}) \right]$ and $\varepsilon_{NW} = E_{NW}/M_{Max} H_{eff}$. According to Equation (8), the magnetization component aligned with the applied magnetic field is represented by $M = M_{Max} \cos(\varphi)$. Utilizing the principles of minimum energy, it is established the conditions $d\varepsilon_{NW}/d\varphi|_{\varphi=\varphi_0} = 0$ and $d^2\varepsilon_{NW}/d\varphi^2|_{\varphi=\varphi_0} \geq 0$ [20].

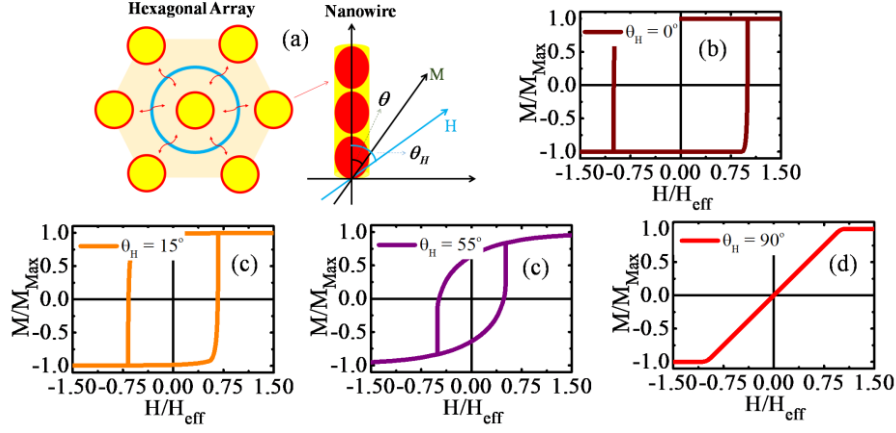

**Figure 1**. (Color online) **(a)** An illustration of a nanowire, characterized by its significantly greater length compared to its diameter, constructed from a continuous chain of interacting grains. Panels **(b)**, **(c)**, **(d)**, and **(e)** present the corresponding hysteresis curves for the described nanowire, highlighting its intriguing magnetic properties and behavior.

The magnetic interactions in nanostructured systems predominantly exhibit PID behavior [25-32], a finding that underscores their complexity and significance. Typically, the relationship between the change in magnetization, Δm, and the reduced magnetization, $m_r$, for systems demonstrating PID behavior is elegantly represented as: $\Delta m_i = A_i[m_r^i(m_r^i - 1)]$, where $A_i$ are constants and $m_r^i$ lies within the closed interval [0, 1]. For this analysis, it is focused on the applied magnetic field directed along the easy axis of magnetization ($\theta_H = 0$), assuming identical grains, with $A_1 = A_2 = A_3 = ... = 1$ and $m_r^i$ also constrained within [0, 1]. **Figure 2(a)** compellingly illustrates the Δm curves, accentuated by a hatched area that delineates the interaction region, which it is defined as interactive maps. This investigation revealed a maximum intensity of demagnetizing interactions at $I_{PID}^{NW}(0) = 0.167$. An in-depth analysis of the angular dependence of these interactions is essential. As highlighted in reference 32, the experimental angular dependence of the Δm curves for PID closely follows a cosine function, reinforcing the predictability of these systems. Thus, one can express this relationship as:

$$\Delta m_i = \sum_{\theta=0}^{\pi/2} B_i(m_d, m_r)\cos(\theta_H). \qquad (9)$$

In this context, one can express $B_i(m_d, m_r)$ in the following manner $A_i[m_r^i(m_r^i - 1)]$. Consequently, equation (9) can be reformulated as:

$$\Delta m_i = \sum_{\theta=0}^{\pi/2}[m_r^i(m_r^i - 1)]\cos(\theta_H). \qquad (10)$$

**Figures 2(a)-(i)** illustrate the intricate maps of magnetic interactions as a function of the applied magnetic field angle, $\theta_H$, determined through equations (1), (2), and (10). Remarkably, these interactions weaken considerably as the magnetic field aligns parallel to the hard axis of magnetization in the nanowire. This trend underscores the profound effect of field orientation, revealing that grain interactions become notably weaker when the magnetic field is perpendicular to the nanowire. Furthermore, the study of domain walls and the propagation of spin waves in the inter-grain regions has shown significant promise [2, 5, 8, 20, 23]. These inter-grain domain walls are not just noteworthy; they are crucial for the exploration of various advanced systems, including skyrmions [33]. This is primarily due to the well-defined nature of the domain walls in these inter-grain regions, which opens up compelling avenues for future research and applications in magnetic materials.

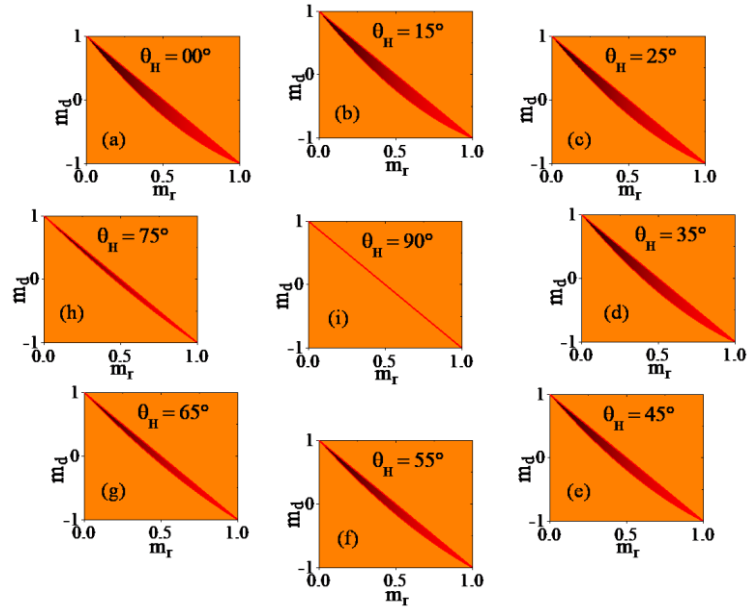

**Figure 2(a)-(i).** (Color online) Interaction maps were meticulously derived from the magnetic interactions as they relate to the applied magnetic field angle, $\theta_H$. These compelling interaction maps emerged from the strategic application of equations (1), (2), and (10), showcasing the robust relationship between magnetic field orientation and its effects.

In **Figure 3(a)**, it is presented a compelling analysis of the angular dependence of coercivity with respect to the applied magnetic field angle, $\theta_H$. This dependence is indicative of the coherent mode of magnetization reversal within the nanowire [1-19].

Notably, the substantial ratio of coercivity to the effective magnetic field underscores the significance of the nanowire's length in this phenomenon. These insights are consistent with the findings illustrated in **Figure 1(b)-(e)**. By employing equations (6) and (10), it is meticulously calculated the intensity values of the PID behaviors across various interaction maps illustrated in **Figure 2**, as shown in **Figure 3(b)**. The data clearly demonstrate that the intensity of magnetic interactions is directly proportional to the change in magnetization, expressed as $I(\theta_H) \propto \Delta m$. For the nanowire under investigation, it is derived the angular dependence of the intensity of magnetic interactions as $I_{PID}^{NW} = I_{PID}^{NW}(0)\cos(\theta_H)$. The excellent fit of numerical data shown in **Figure 3** validates this model, revealing that $I_{PID}^{NW}(0)$ is precisely 0.167. This robust relationship not only reinforces our understanding of magnetic interactions in nanowires but also highlights their potential applications in advanced magnetic devices.

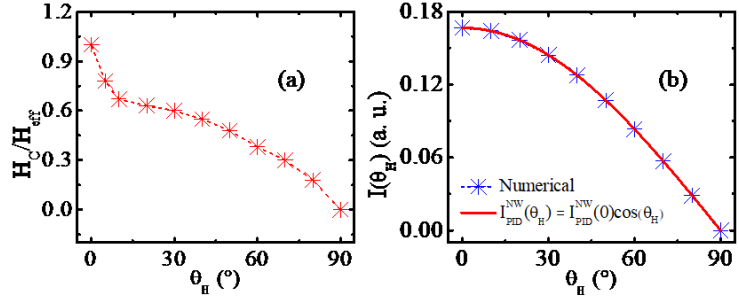

**Figure 3**. (Color online) Data for a nanowire reveals important insights: **(a)** Angular dependency of coercivity demonstrates how this property varies with direction. **(b)** Angular variation of the intensity value $I_{PID}^{NW}(\theta_H)$, extracted from the detailed maps presented in **Figure 2**. This relationship was accurately modeled using the equation $I_{PID}^{NW} = I_{PID}^{NW}(0)\cos(\theta_H)$ $I_{PID}^{NW}(0) = 0.167$. These findings underscore the critical influence of angular orientation on nanowire performance.

**b) 3D, comprising a hexagonal array of nanowires**

The physical properties of three-dimensional materials differ from those of one-dimensional materials, such as a single nanowire. An array of nanowires can be composed of metals, semiconductors, and magnetic materials, exhibiting interactions unique to their morphology [1-8]. When considering the magnetic energy density of a hexagonal array of nanowires, the total free energy density can be expressed as follows:

$$E_{NWA} = \pi M_{Max}^2 \left[\left(\frac{3V_e k_n}{\pi r^3}\right) + (1 - 3N_{/\!/}) - 3P\right] \sin^2(\varphi + \theta_H) - M_{Max} H\cos(\varphi) - A\cos(\varphi + \theta_H). \tag{11}$$

In this equation, the difference between equations (7) and (11) represents the energy densities associated with the packing of the nanowires in the array, given by $E_{PID}^{NWA} = -3\pi M_{Max}^2 \sin^2(\varphi + \theta_H)$, and the exchange energy, denoted as $E_{PIM}^{Exc} = -A\cos(\varphi + \theta_H)$ [2, 5, 8, 20, 23], where P is the packing factor of the nanowires in the array, and A is the exchange constant. Equation (11) can be rewritten as:

$$\varepsilon_{NWA} = \frac{1}{2}\sin^2(\varphi + \theta_H) - h\cos(\varphi) - g\cos(\varphi + \theta_H). \tag{12}$$

Here, the factors $h = H/H'_{eff}$ and $g = A/M_{Max} H'_{eff}$ incorporate the effective field term $H'_{eff} = 2\pi M_{Max}\left[1 + 3(V_e k_n/\pi r^3 - N_{/\!/} - P)\right]$, and $\varepsilon_{NWA} = E_{NWA}/M_{Max} H'_{eff}$. The hysteresis curves calculated using the minimum energy conditions discussed previously are shown in **Figures 4(a)-(c)**. The energy densities associated with the packing of the nanowires in the array and the exchange produce packing-induced dynamics (PID) and packing-induced magnetization (PIM) behaviors, respectively. The exchange interactions are primarily influenced by the defects arising from the packing of the nanowires within the array. In this context, it is essential to consider the defect term $D_i(m_d, m_r)\sin(\theta_H)$ in equation (9) to derive the magnetic interactions for a hexagonal array of nanowires:

$$\Delta m_i = \sum_{\theta=0}^{\pi/2}[B_i(m_d, m_r)\cos(\theta_H) + D_i(m_d, m_r)\sin(\theta_H)]. \tag{13}$$

The term $D_i(m_d, m_r)$ is defined by the expression $-m_r^2(m_r - 1)/2$, representing the average of the PID and PIM interactions within the array. Therefore, equation (13) is rewritten as:

$$\Delta m_i = \sum_{\theta=0}^{\pi/2}[m_r^i(m_r^i - 1)]\cos(\theta_H) - \left[\frac{(m_r^i)^2}{2}(m_r^i - 1)\right]\sin(\theta_H). \tag{14}$$

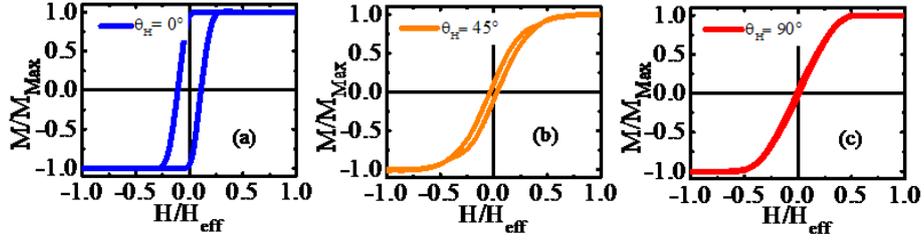

**Figure 4**. (Color online) Panels **(a)**, **(b)**, and **(c)** present the compelling hysteresis curves corresponding to the hexagonal array of nanowire, highlighting its unique properties and performance characteristics.

**Figure 5(a)-(i)** presents compelling maps of magnetic interactions at various angles of the applied magnetic field, meticulously calculated using equations (1), (2), and (14). Consistent with observations in single nanowires, the magnetic interactions diminish as the applied magnetic field aligns with the hard axis of magnetization in the hexagonal array of nanowires. Notably, the PID exhibits a sharp decline in response to the applied magnetic field, while the PIM simultaneously rises. This intriguing behavior arises from the unique coupling between spins within the three-dimensional nanostructure, which plays a crucial role in the excitations generated by dipolar and exchange spin-waves. The insights derived from this study underscore the complex interactions at play and their significance in understanding magnetic behavior in nanostructured systems.

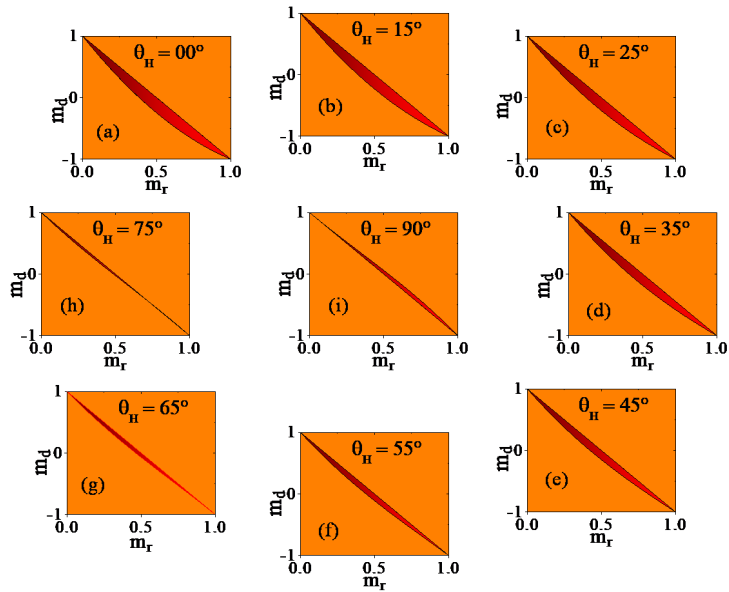

**Figure 5(a)-(i)**. (Color online) Interaction maps effectively illustrate the magnetic interactions across a range of angles for the applied magnetic field. These compelling

results were derived from equations (1), (2), and (10), highlighting the significance of our analytical approach.

It is crucial to account for the energy density arising from the arrangement of nanowires within the array when conducting a comprehensive analytical analysis of the angular variation based on the intensity derived from the PID behavior. The intensity of the PID behavior for a hexagonal array of nanowires can be expressed by the following equation:

$$I_{PID} = I_{PID}^{NW}(0)\cos(\theta_H) + I_{PID}^{NWA}\sin^2(\varphi + \theta_H). \quad (15)$$

Conversely, the angular variation of the intensity associated with the PIM behavior is represented by the equation:

$$I_{PIM} = I_{PIM}^{NWA}(0)\cos(\varphi + \theta_H). \quad (16)$$

These equations (15) and (16) provide a robust framework for independently assessing the intensity of interactions, a factor critically relevant to the propagation of spin waves [8, 34]. In **Figure 6(a)**, it is presented the angular dependence of coercivity as a function of the applied magnetic field angle, $\theta_H$. This relationship illustrates the various modes of magnetization reversal occurring in the hexagonal array of nanowires. Notably, the small ratio of coercivity to the effective magnetic field underscores the unique three-dimensional structure of the array, a finding that aligns with the insights provided in **Figures 4(b)-(d)**. By employing equations (6) and (14), it is calculated the PID and PIM behaviors for the distinct interaction maps showcased in **Figure 5**, as depicted in **Figure 6(b)**. This comprehensive analysis not only strengthens our understanding of the system but also highlights the significant implications for future research in the field.

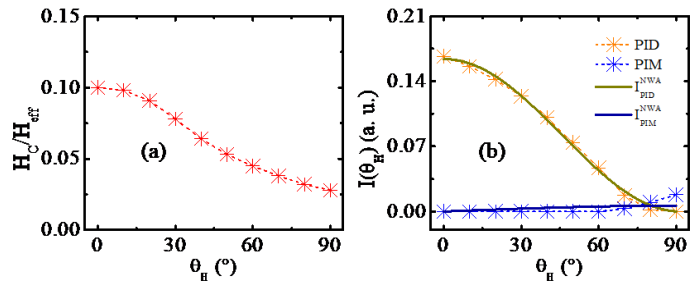

**Figure 6**. (Color online) Insights into the hexagonal array of nanowires reveal crucial data: **(a)** Analyzing the angular dependence of coercivity demonstrates its significant variation with angle. **(b)** Angular profile of the intensity value $I_{PID}^{NW}(\theta_H)$, extracted from the detailed maps in **Figure 5**, underscores the intricate relationships within the system, emphasizing the robustness of our findings.

**c) Analyzing the minimum free energies in different dimensions**

**Figure 7** compellingly illustrates the behavior of free energy densities for both a single nanowire and a hexagonal array of nanowires. In **Figure 7(a)**, it is presented critical conditions leading to energy minima, and the inset reveals how energy fluctuates with respect to the direction of the applied magnetic field, specifically at $f = 1$, and $\varphi = -90^0$. Notably, at $\varphi = -90^0$ and $\theta_H = 90^0$, the free energy not only achieves but decisively reaches its minimum value. **Figure 7(b)** further delineates the varying conditions for energy minima based on the orientation of the applied magnetic field. The inset confirms that when $h = 1$, $g = 1$, and $\varphi = -90^0$, the free energy is optimized when $\theta_H = 90^0$. Moreover, **Figure 7** underscores the vital equilibrium condition of magnetization during magnetic interaction measurements, illuminating the exciting potential for the generation of spin waves during equilibrium precession.

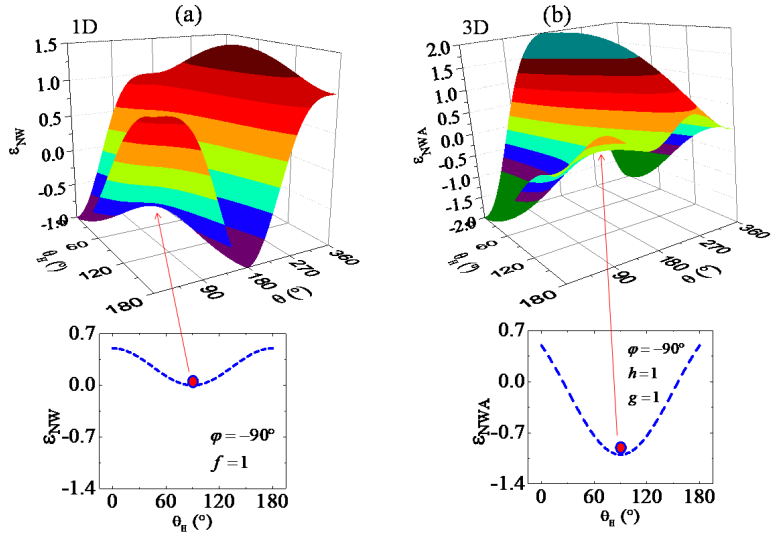

**Figure 7.** (Color online). **(a)** Findings compellingly demonstrate that the positions of minimum energy are strategically located along the axis of the nanowire. The inset in this figure effectively highlights the energy variation in relation to the direction of the applied magnetic field, specifically at $f = 1$ and $\varphi = -90^0$. **(b)** Furthermore, the analysis reveals that the hexagonal array of nanowires exhibits the same advantageous characteristic, with minimum energy positions again aligning along the nanowires' axis. The accompanying inset illustrates how energy varies with the applied magnetic field direction for the hexagonal array parameters $h = 1$, $g = 1$, and $\varphi = -90^0$. Remarkably, in both cases, the energy reaches its absolute minimum at $\theta_H = 90º$, reinforcing the consistency of this phenomenon.

## IV. CONCLUSION

According to the magnetic domain mechanism, nanoparticles serve as exemplary models of single domain structures. Conversely, grains arranged in a linear fashion create a network of dipole moments. This makes nanowires an outstanding example of organized nanostructures, further enhancing their potential for innovative applications within the systems discussed. Moreover, this study not only presents a compelling opportunity to calculate values and analyze the intensity of interactions in magnetic nanostructures, but it also compellingly illustrates how these interactions significantly influence spin-wave excitations. From the perspective of fundamental physics, our results mark a crucial advancement in understanding the dynamics of interactions within nanostructures. Additionally, this proposal holds the promise of accelerating the development of cutting-edge spintronic devices that leverage inter-grain interactions, paving the way for revolutionary technological advancements.

**Acknowledgements**

This research was supported by Conselho Nacional de Desenvolvimento Científico e Tecnológico (CNPq) with Grant Number: 309982/2021-9, Coordenação de Aperfeiçoamento de Pessoal de Nível Superior (CAPES) with Grant Number: PROAP2024UFRPE, and Fundação de Amparo à Ciência e Tecnologia do Estado de Pernambuco (FACEPE) with Grant Number: APQ-1397-3.04/24.

**Contributions**

All results were obtained by J. H.

**Conflict of interest**

The author declares that they have no conflict of interest.

**Data availability statement**

Data will be made available on reasonable request.